\documentclass[aps,preprint,showpacs, superscriptaddress,onecolumn, floatfix, nofootinbib]{revtex4}%
\usepackage{amsfonts}
\usepackage{amsmath}
\usepackage{amssymb}
\usepackage{graphicx}
\usepackage{subfigure}
\usepackage[svgnames]{xcolor}%
\begin{document}
\preprint{JLAB-THY-07-701}

\title{$1/N_{c}$ Expansion in QCD:
Double-Line Counting Rules and the Undeservingly Discarded $U(1)$ Ghost\footnotetext{Notice: Authored by Jefferson Science Associates, LLC under U.S. DOE Contract No. DE-AC05-06OR23177. The U.S. Government retains a non-exclusive, paid-up, irrevocable, world-wide license to publish or reproduce this manuscript for U.S. Government purposes.}}
\author{Hrayr H. Matevosyan}
 \affiliation{Thomas Jefferson National Accelerator Facility, 12000
              Jefferson Ave., Newport News, VA 23606, USA}
\affiliation{Louisiana State University, Department of Physics \& Astronomy, 202 Nicholson
Hall, Tower Dr., Baton Rouge, LA 70803, USA}
\author{Anthony W. Thomas}
\affiliation{Thomas Jefferson National Accelerator Facility, 12000 Jefferson Ave., Newport
News, VA 23606, USA}
\affiliation{College of William and Mary, Williamsburg VA 23187, USA}

\begin{abstract}
The $1/N_{c}$ expansion is one of the very few methods we have for generating a systematic expansion of QCD at the energy scale relevant to hadron structure. The present formulation of this theory relies on the double-line notation for calculating the leading order of a diagram in the $1/N_{c}$  expansion, where the local $SU(N_{c})$ gauge symmetry is substituted by a $U(N_{c})$ symmetry and the associated $U(1)$ ghost field is ignored. In the current work we demonstrate the insufficiency of this formulation for describing certain non-planar diagrams. We derive a more complete set of Feynman rules that include the $U(1)$ ghost field and provide a useful tool for calculating both color factors and $1/N_{c}$ orders of all color-singlet diagrams. 

\end{abstract}

\date{\today}
\maketitle

\section{Introduction}

 Quantum Chromodynamics in the energy region relevant to hadron structure does not have a small parameter that can be used in an expansion. Gerard 't Hooft proposed to generalize QCD to an arbitrary number of colors, $N_{c}$, and to use the inverse of the number of colors, $1/N_{c}$, as such an expansion parameter.  For the expansion to be convergent, the number of the colors is assumed to be very large, while $g^{2}/N_{c}$ is kept fixed \cite{'tHooft:1973jz}. This model has
proven to be a very useful analytical tool for obtaining exact results in
QCD-like theories and hadron phenomenology \cite{'tHooft:1974hx}, \cite%
{Witten:1979kh}, \cite{Manohar:1998xv}. In this formalism the double-line counting scheme gives diagrammatic rules for ordering the Feynman diagrams according to their leading power of $1/N_{c}$. These rules can be used to prove, on topological grounds, that the diagrams that can be drawn on a plane without any of the lines crossing (planar diagrams) are of leading order with respect to those which can only be drawn with some of the lines crossing (non-planar diagrams).

\begin{figure}
[ptbh]
\begin{center}
\includegraphics[width=0.8\textwidth
]%
{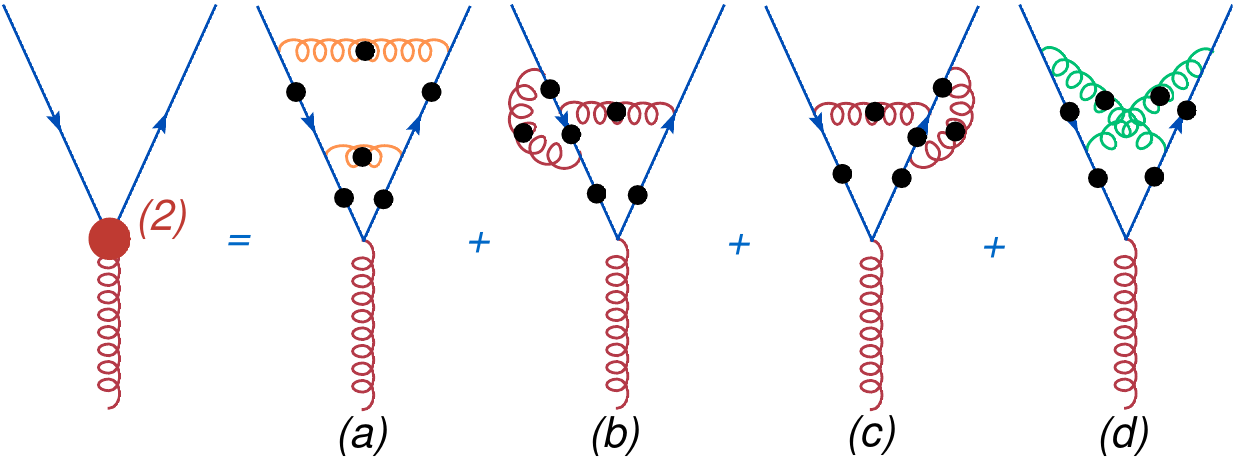}%
\caption{Vertex diagrams at $\mathcal{O}(g^{5})$. Here the large red circle
denotes the 2-nd order dressed quark-gluon vertex function ( 2 in the
parenthesis denotes the number of gluon lines contributing to the vertex) and
the small black circles on the propagators denote that the propagators are
fully dressed. The non-planar diagram (d) proves to have significantly larger color factor than the planar diagrams (a)-(c).}%
\label{PLOT_FV_ORDER_2}%
\end{center}
\end{figure}

 As a simple, introductory example we recall our recent work exploring the consequences of using different models for dressing quark-gluon vertex
diagrams for the solutions of the quark Gap equation \cite{Matevosyan:2007cx}. There we came across the diagrams depicted in Fig. \ref{PLOT_FV_ORDER_2}, contributing at order $\mathcal{O}(g^{5})$ in dressing the vertex. A naive interpretation of the planarity arguments leads to the conclusion that the diagram in Fig.  \ref{PLOT_FV_ORDER_2}d should have a smaller color factor than those depicted in Figs.  \ref{PLOT_FV_ORDER_2}a-c, since they all have same number of factors of $g$.   

On the other hand, the color
factors of the diagrams in Fig. \ref{PLOT_FV_ORDER_2} are easy to calculate
explicitly using the properties of the generating matrices of the fundamental
representation of the $SU(N_{c})$ group. The diagrams in Fig. \ref{PLOT_FV_ORDER_2}a-c have the the same color factors:%
\begin{equation}
\text{Cf}_{2}^{\text{Pl}}t^{a}=t^{c}t^{b}t^{a}t^{b}t^{c}=\frac{1}{4N_{c}^{2}}t^{a}.\label{EQ_CF_IMP}%
\end{equation}

The color factor of Fig. \ref{PLOT_FV_ORDER_2}d can also be easily
calculated:%
\begin{equation}
\text{Cf}_{2}^{\text{NPl}}t^{a}=t^{c}t^{b}t^{a}t^{c}t^{b}=\frac{N_{c}^{2}%
+1}{4N_{c}^{2}}t^{a}. \label{EQ_CF_NPL}%
\end{equation}

\begin{figure}[ptb]
\centering \subfigure[] {
\includegraphics[width=0.5\textwidth]{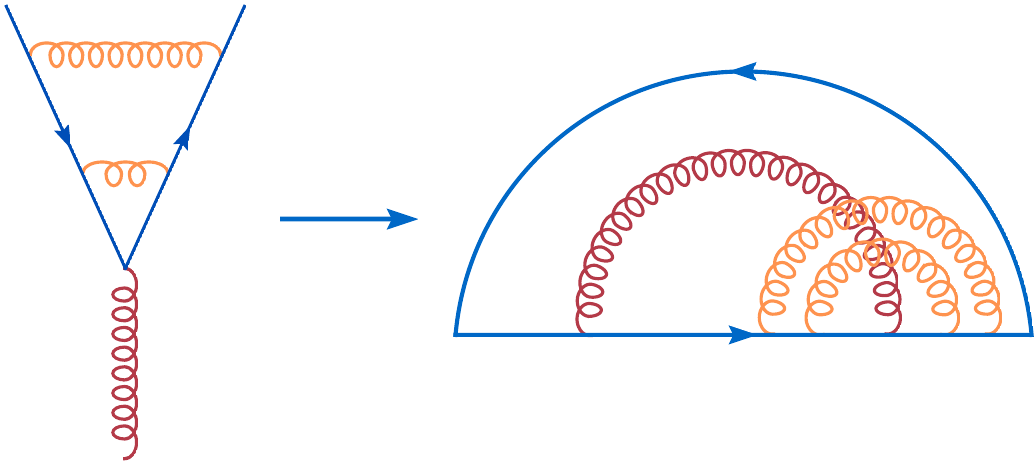} } \vspace{1cm} \subfigure[] {
\includegraphics[width=0.5\textwidth]{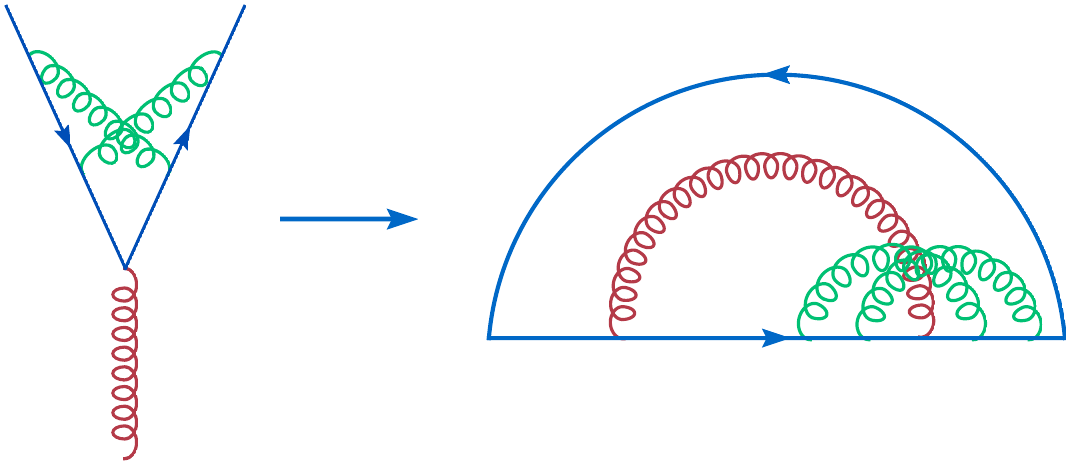}
}\caption{The non-planar diagrams obtained from inserting the corresponding
vertices into the quark self-energy term and closing the fermion loop to form
a color-singlet object. This shows that there is no clear indication as to
which diagram is dominant in a $1/N_{c}$ expansion.}%
\label{PLOT_VERT_SEN_2AD}%
\end{figure}

This result seems to be in contradiction with the planarity arguments
associated with the $1/N_{c}$ expansion. The qualitative solution of this
puzzle can be understood if we look more closely at the topological criteria
in $1/N_{c}$. The criterion of planarity can be applied only to color-singlet
objects if it is to be unambiguous. Moreover, the outer edge of the diagram must
be a quark loop\footnote{It can be proven by invoking the generalized polyhedral ($\mathrm{Poincar\acute{e}}$) formula that in $U(N_{c})$ theory the leading $N_{c}$ order of a particular connected vacuum (color-singlet) diagram can be expressed as $\chi=2-2h-b$, where $h$ is the number of handles and $b$ is the number of boundaries of the manifold on which the diagram is drawn.  Each quark loop represents a boundary, thus a graph with one quark loop should be drawn on a surface with a topology of a sphere with one hole (a circle bounded by the quark line) versus a graph with no quarks which is drawn on a surface of a sphere (infinite plane). For planar graphs $h=0$, where for non-planars one needs to insert handles in the surface for the intersecting gluon lines to jump over each other.} \cite{Manohar:1998xv}. The most logical way of constructing a
color-singlet object from the vertex diagrams of Fig. \ref{PLOT_FV_ORDER_2}
is by inserting them into the quark self-energy term of the Gap\ equation and
closing the fermion loop in a manner that will leave the quark loop on the
outer edge of the diagrams. This is demonstrated in Fig.
\ref{PLOT_VERT_SEN_2AD} for the diagrams shown in Figs. \ref{PLOT_FV_ORDER_2}a and
\ref{PLOT_FV_ORDER_2}d.

This demonstrates that both vertex diagrams yield non-planar diagrams, which
need to be considered in more detail in a $1/N_{c}$ expansion in order to
determine their leading order contribution. The $1/N_{c}^{2}$ suppression of the diagrams in Fig. \ref{PLOT_FV_ORDER_2}a-c, with respect to \ref{PLOT_FV_ORDER_2}d, ought to be understood by applying double-line counting rules, but it is easy to convince oneself that the double-line rules yield exactly the same order in $1/N_{c}$ expansion for all diagrams. In the next section we investigate this problem in detail and formulate a revised set of double-line rules for both calculating color factors and for counting the $1/N_{c}$ order of color-singlet Feynman diagrams.

\section{Revising Double-Line Counting Rules}

A careful examination of the $1/N_{c}$ power behavior of the non-planar
diagrams requires a short review of the introduction to the double-line counting model of
the $1/N_{c}$ expansion (also known as large $N_{c}$ QCD) \cite{Manohar:1998xv}.
As we mentioned in the previous section, QCD in the low energy region, where
most of the hadrons lie, does not have an explicit parameter in which it can
be expanded. 't Hooft's proposition was to use $1/N_{c}$, assuming $N_{c}$ to
be so large that the expansion will be sensible. In the discussion from now on
we will drop the $c$ index from the number of colors, writing it simply as $N$.

The expansion is setup by extending QCD to $N$ colors and $N_{f}$ flavors of
quarks. The underlying local gauge symmetry is $SU(N)$ in the fundamental
representation. Here the color parts of the quark fields are represented by
$N$ dimensional vectors and the gauge fields are represented by Hermitian
$N\times N$ traceless matrices $t^{a}$, $A_{\mu}\equiv t^{a}A_{\mu}^{a}$, with
index $a\in\left\{  1,2,..,N^{2}-1\right\}  $ in the adjoint representation of
$SU(N)$. In the $1/N$ expansion the gauge coupling constant $g$ is taken to be
$g=g_{0}/\sqrt{N}$, which is necessary to keep the theory sensible as
$N\rightarrow\infty$.

The corresponding Lagrangian density is:%
\begin{equation}
\mathcal{L}_{QCD}=\sum_{f=1}^{N_{f}}\bar{\psi}_{f}(i\gamma^{\mu}D_{\mu}%
-m)\psi_{f}-\frac{1}{2}Tr\left[  F^{\mu\nu}F_{\mu\nu}\right]  ,
\label{EQ_COL_LAGRANGIAN}%
\end{equation}
with the covariant derivative defined via gauge field:%
\begin{equation}
D_{\mu}\equiv\partial_{\mu}+i\frac{g_{0}}{\sqrt{N}}A_{\mu},
\label{EQ_COL_COV_DER}%
\end{equation}
and the field strength as:%
\begin{equation}
F_{\mu\nu}^{a}\equiv\partial_{\mu}A_{\nu}^{a}-\partial_{\nu}A_{\mu}^{a}%
+i\frac{g_{0}}{\sqrt{N}}f_{abc}\left[  A_{\mu}^{b},A_{\nu}^{c}\right]  ,
\label{EQ_COL_FIELD_STRENGTH}%
\end{equation}
From now on we will ignore the flavor index of the quarks, as it will be
irrelevant to our discussion. We also note that the large $N$ limit can be
taken with either $N_{f}$ or $N_{f}/N$ fixed.

\begin{figure}[ptb]
\centering\subfigure[]{
	\includegraphics[width=0.5\textwidth]{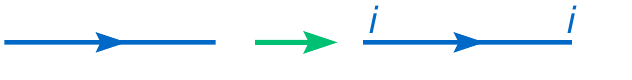}} \vspace{1cm}
\subfigure[]{
	\includegraphics[width=0.7\textwidth]{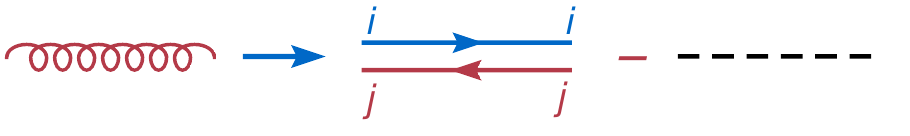}
}\caption{The double-line notation for the quark and gluon propagators in a 
$1/N_{c}$ expansion.}%
\label{PLOT_PROPS_DL}%
\end{figure}

The counting of the color factors of the Feynman diagrams is necessary in
assessing their importance in this counting scheme. 't Hooft proposed
a\ diagrammatic method for counting the color factors of the diagrams, the
so-called double-line notation. The method is easy to understand by
considering the color structure of the theory's Green's functions. The quark
propagator has the following structure%
\begin{equation}
\left\langle 0\right\vert \psi^{i}(x)\bar{\psi}^{j}(y)\left\vert
0\right\rangle =\delta^{ij}S\left(  x-y\right)  ,\label{EQ_COL_QUARK_PROP}%
\end{equation}
where the color indices $i,j\in\left\{  1,2,...,N\right\}  $. Thus, the
propagator is depicted diagrammatically by an arrowed line with the same color
indices on both ends because of $\delta^{ij}$, see Fig. \ref{PLOT_PROPS_DL}a.

The gauge field propagator is%
\begin{equation}
\left\langle 0\left\vert A_{\mu}^{a}(x)A_{\nu}^{b}(y)\right\vert
0\right\rangle =\delta^{ab}D_{\mu\nu}\left(  x-y\right).
\label{EQ_COL_GAUGE_PROP}%
\end{equation}

Using the following $SU(N)$ identity \cite{Cvitanovic:1976am}%
\begin{equation}
\left(  t^{a}\right)  _{j}^{i}\left(  t^{a}\right)  _{l}^{k}=\frac{1}{2}%
\delta_{l}^{i}\delta_{j}^{k}-\frac{1}{2N}\delta_{j}^{i}\delta_{l}^{k},
\label{EQ_COL_SUN_REL}%
\end{equation}
and the normalization condition of the $t^{a}$ matrices, the color indices of
the propagator (\ref{EQ_COL_GAUGE_PROP}) can be written explicitly%
\begin{equation}
\left\langle 0\left\vert \left(  A_{\mu}^{a}(x)\right)  _{j}^{i}\left(
A_{\nu}^{b}(y)\right)  _{l}^{k}\right\vert 0\right\rangle =\left(  \frac{1}%
{2}\delta_{l}^{i}\delta_{j}^{k}-\frac{1}{2N}\delta_{j}^{i}\delta_{l}%
^{k}\right)  D_{\mu\nu}\left(  x-y\right).  \label{EQ_COL_GAUGE_SUN_PROP}%
\end{equation}

The $U(N)$ identity corresponding to (\ref{EQ_COL_SUN_REL}) has the
following form%
\begin{equation}
\left(  t^{a}\right)  _{j}^{i}\left(  t^{a}\right)  _{l}^{k}=\frac{1}{2}%
\delta_{l}^{i}\delta_{j}^{k},\label{EQ_COL_UN_REL}%
\end{equation}
which, in the usual large-N expansion is usually substituted into the gluon propagator. The gluon propagator
is then diagrammatically depicted as a double line, with opposite arrows and
the same color indices on both ends of each line, as implied by the Kronecker
deltas in (\ref{EQ_COL_UN_REL}). The second term on the right hand side (RHS)
of Eq. (\ref{EQ_COL_SUN_REL}) is dropped, assuming it will be unimportant in
the $1/N$ counting, as it is suppressed by a factor of $1/N$ .

Here we propose to keep the second term on the RHS of Eq.
(\ref{EQ_COL_SUN_REL}), which corresponds to a $U(1)$ ghost field, which is
necessary to cancel the extra $U(1)$ gauge boson in relating the $U(N)$ gauge
theory to $SU(N)$. We propose to depict the ghost field as a single dotted
line with no color indices, as it transfers no color because of the Kronecker
deltas that contract the color factors at each end of the line - see Eq.
(\ref{EQ_COL_SUN_REL}). The corresponding diagrams are depicted in Fig.
\ref{PLOT_PROPS_DL}b, where the double-line object corresponds to the $U(N)$
gluon propagator, and the ghost field has a negative sign as one can see from
(\ref{EQ_COL_GAUGE_SUN_PROP}). In counting the powers of $N$, each ghost field
will enter with a $1/N$ factor which also arises from the form of the propagator.
This extra penalty of $1/N$ is the reason the ghosts are generally ignored in
the literature. We also note that with each double-line $U(N)$ gluon and
$U(1)$ ghost propagator, we have to include a factor of $1/2$ to account for
the proper normalization.

\begin{figure}[ptb]
\includegraphics[width=0.8\textwidth]{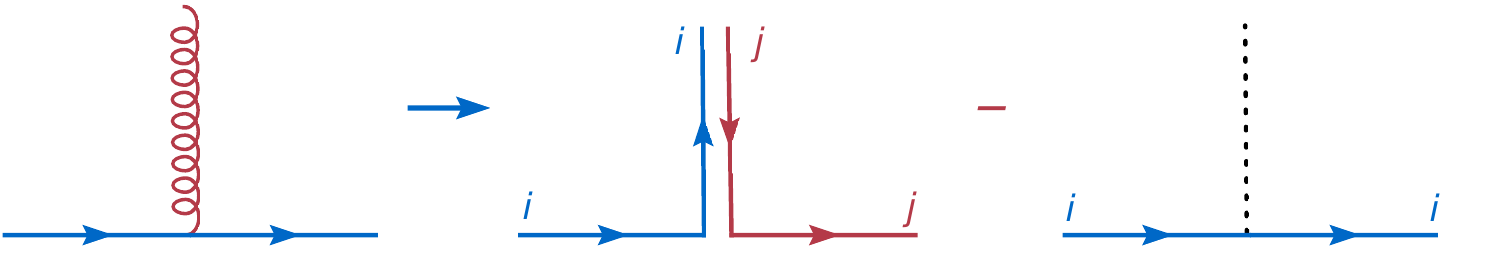}
\caption{The double-line notation for the quark-gluon vertex function in a 
$1/N_{c}$ expansion.}%
\label{PLOT_VERTS_DL}%
\end{figure}

The quark-gluon vertex is generated by the following term of the Lagrangian
\begin{equation}
g\bar{\psi_{i}} (A^{a})_{j}^{i}\psi^{j},\label{EQ_COL_QG}%
\end{equation}%
Thus the quark-gluon vertex can be depicted in the
double-line and dotted notation as shown in Fig. \ref{PLOT_VERTS_DL}. We note that there is no $U(N)$ gluon-ghost interaction, as these fields are commutative. We also note here, that in the general literature on the $1/N$ expansion, the ghost contribution is commonly ignored. 

The color structure of the gluon 3-point function is given by the totally antisymmetric structure with the color indices of the gluon legs $f^{abc}$. As noted in the Ref. \cite{Cvitanovic:1976am}, a simple $SU(N)$ relation allows one to reduce the problem to the case of quark-gluon vertices, which we have already discussed:
\begin{equation}
f^{abc}=2\times Tr[t^{a}t^{b}t^{c}-t^{c}t^{b}t^{a}].\label{EQ_COL_3G}%
\end{equation}%

\begin{figure}[ptb]
\includegraphics[width=0.6\textwidth]{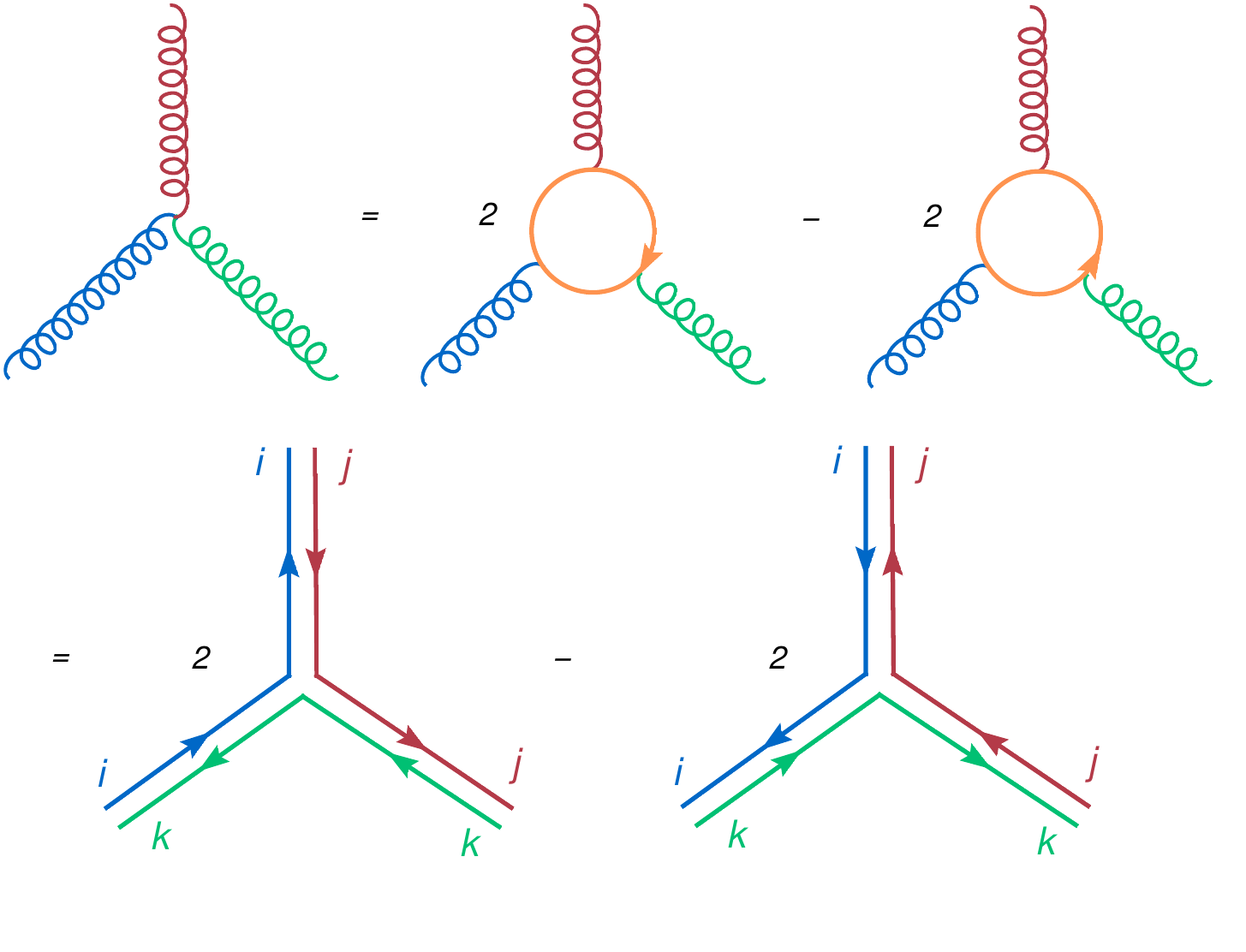} 
\caption{The double-line notation for the 3-gluon vertex function in a 
$1/N_{c}$ expansion.
}%
\label{PLOT_3G_VERT_DL}%
\end{figure}

 The diagrammatic notation is shown in Fig. \ref{PLOT_3G_VERT_DL}, where the diagrams containing ghost lines cancel each other.

\begin{figure}[ptb]
\includegraphics[width=0.6\textwidth]{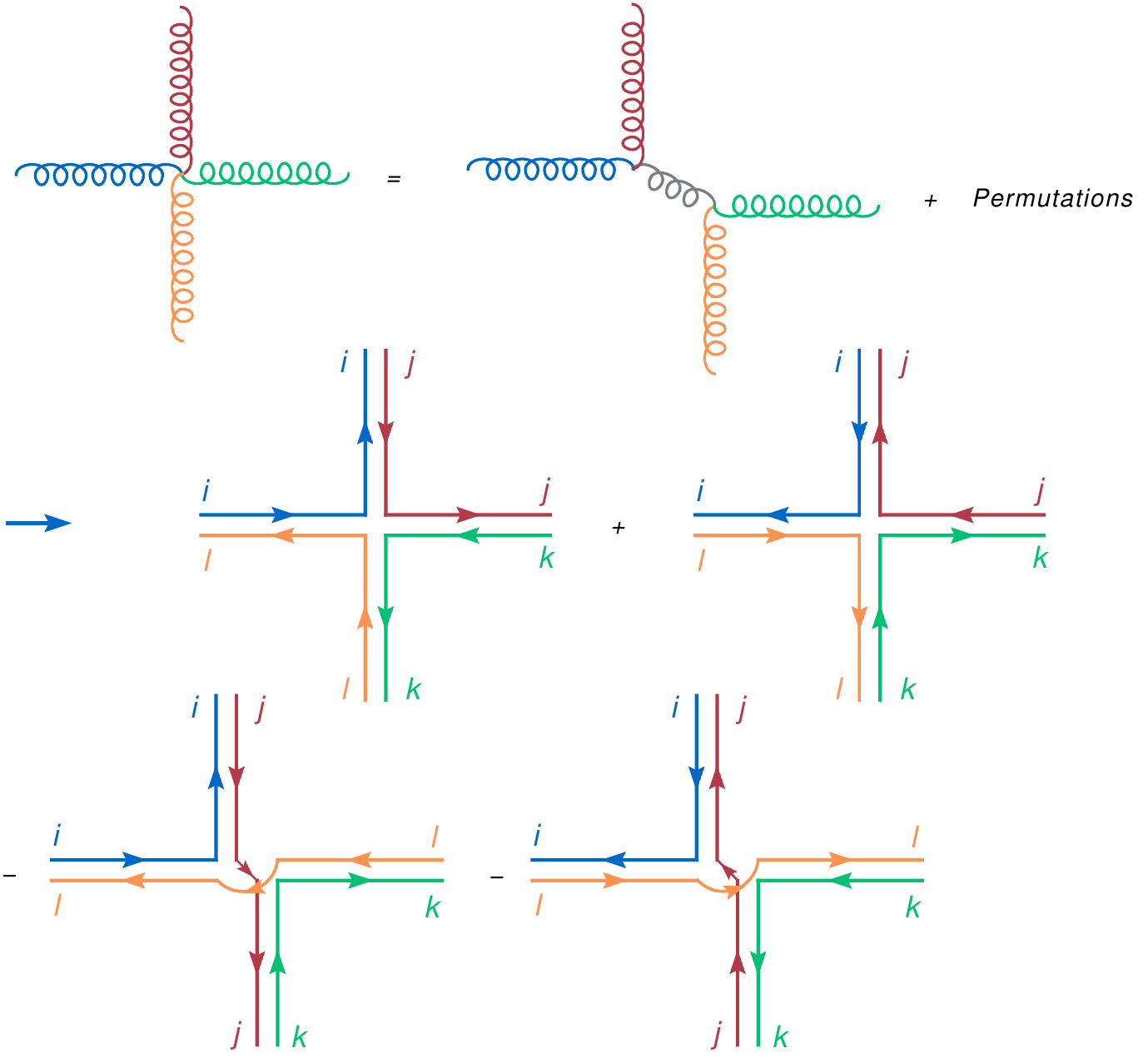} 
\caption{The double-line notation for the 4-gluon vertex function in a 
$1/N_{c}$ expansion.}%
\label{PLOT_4G_VERT_DL}%
\end{figure}

The gluon 4-point function has three parts, where all possible pairs of gluon legs are formed, with appropriate color and Dirac factors. A typical color structure is:
\begin{equation}
f^{abe}f^{ecd},%
\end{equation}%
which can be diagrammatically depicted as an exchange of a single gluon between the four gluon legs as depicted in Fig.  \ref{PLOT_4G_VERT_DL}. Then each 3-gluon vertex can be expanded using the rules in Fig. \ref{PLOT_3G_VERT_DL}. The double-line diagram expansion is shown in Fig. \ref{PLOT_4G_VERT_DL}, without the multiplicative factors $4$ arising form relation (\ref{EQ_COL_3G}). The $1/N$ expansion for Faddeev-Popov ghost propagator and the gluon-ghost-ghost vertex is completely analogous to that for the gluon propagator and the 3-gluon vertex, and we therefore also skip their treatment here.

With this diagrammatic notation, the color factor of a diagram may be counted by
replacing the Green's functions with the corresponding double lined diagrams
and counting the number of closed color index loops, each accounting for a
factor of $N$. The $1/N$ counting is unambiguous only for color-singlet
objects, where the corresponding diagram will not have any open color index
lines. The overall order of the diagram is then calculated by taking into
account the number of $1/\sqrt{N}$ factors from the gauge coupling constant
and the $1/N$ factors from the ghost propagators. That is, we write the order, $O$, as:%
\begin{equation}
O=N^{L-C/2-G}, \label{EQ_LARGEN_LG}%
\end{equation}
where $L$ is the number of closed color index loops, $C$\ is the number of
couplings, $g$, and $G$\ is the number of ghost propagators in the diagram.

We note that a similar scheme for calculating $SU(N)$ color factors was developed earlier in the literature (the so-called bird-track notation) - see Ref. \cite{Cvitanovic:1976am} and references within. This was where we learned how to implement the color factors for 3- and 4-point gluon functions. On the other hand, our version of the Feynman rules gives a more intuitive method for calculating the color factors by preserving the topological structure of the original diagram and keeping track of numerical factors arising from ghost lines, leaving less space for confusion when dealing with non-trivial diagrams with large number of quark and gluon lines.

\begin{figure}
[ptb]
\begin{center}
\includegraphics[width=0.6 \textwidth
]%
{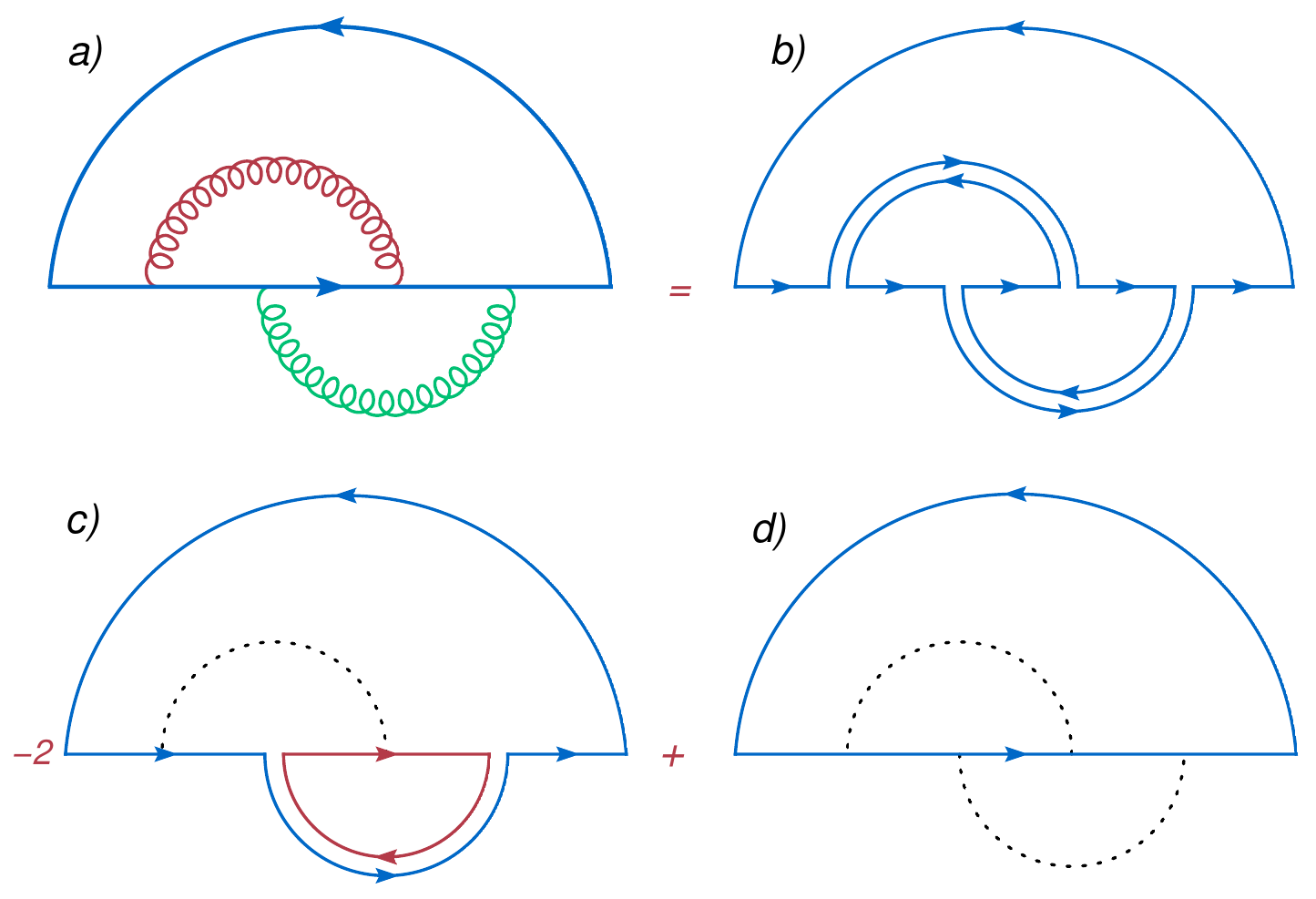}%
\caption{The double-line analysis of the lowest order non-planar diagram with
quark loop, where the $U(1)$ ghosts were also included.}%
\label{PLOT_SELF_EN_1_DL}%
\end{center}
\end{figure}

These rules are easy to demonstrate on the simple example depicted in Fig.
\ref{PLOT_SELF_EN_1_DL}. The color factor and $1/N$ order of the diagram, Fig. \ref{PLOT_SELF_EN_1_DL}a is easy to calculate%

\begin{align}
\text{Cf}_{1}^{a)}  &  =Tr[t^{a}t^{b}t^{a}t^{b}]=-\frac{N^{2}-1}{4N},\\
O_{1}^{a)}  &  =\text{Cf}_{1}^{a)}\times\left(  \frac{1}{\sqrt{N}}\right)
^{4}=-\frac{N^{2}-1}{4N^{3}}.\nonumber
\end{align}

The rules described above allow for an easy evaluation of the color factors
for the diagrams in Figs. \ref{PLOT_SELF_EN_1_DL}b-\ref{PLOT_SELF_EN_1_DL}d:%
\begin{gather}
\text{Cf}_{1}^{b)}=\frac{1}{4}N\\
O_{1}^{b)}=\text{Cf}_{1}^{b)}\times\left(  \frac{1}{\sqrt{N}}\right)
^{4}=\frac{1}{4N}\nonumber\\
\text{Cf}_{1}^{c)}=\frac{1}{4}N^{2}\times\frac{1}{N}=\frac{1}{4}N\nonumber\\
O_{1}^{c)}=\text{Cf}_{1}^{c)}\times\left(  \frac{1}{\sqrt{N}}\right)
^{4}=\frac{1}{4N}\nonumber\\
\text{Cf}_{1}^{d)}=\frac{1}{4}N\times\frac{1}{N^{2}}=\frac{1}{4N}\nonumber\\
O_{1}^{d)}=\text{Cf}_{1}^{d)}\times\left(  \frac{1}{\sqrt{N}}\right)
^{4}=\frac{1}{4N^{3}}\nonumber
\end{gather}

It is easy to check that%
\begin{equation}
\text{Cf}_{1}^{a)}=\text{Cf}_{1}^{b)}-2\text{Cf}_{1}^{c)}+\text{Cf}_{1}^{d)},
\end{equation}
as depicted in the Fig. \ref{PLOT_SELF_EN_1_DL}.

Here we see the importance of the $U(1)$ ghost fields, where the term Cf$_{1}^{c)}$
is equal to Cf$_{1}^{b)}$, but is multiplied by a factor of two with the
opposite sign. In the literature this example is demonstrated without the
ghost fields, which should yield the wrong sign, but accidentally the right
order in the $1/N$ expansion. We want to emphasize that the ghost fields can
be discarded in describing the $SU(N)$ theory, \textit{only }if we consider the
leading order diagrams with the given number of quarks and gluons - that is,
only for planar diagrams. For planar diagrams, each $U(N)$ gluon forms a new
color index loop which cancels the penalty of the $1/N$\ induced by the two
factors of $g$ it carries. On the other hand, a ghost line does not create a
color loop and carries an extra $1/N$ penalty for the propagator, thus
contributing at order $1/N^{2}$ compared to the corresponding $U(N)$ gluon.
However, for the non-planar diagrams, the $U(N)$ gluon may not create, or it may
even destroy, a color index loop, where the ghost stays \textquotedblleft
invisible\textquotedblright\ (same old ghostly habits) to the color loops.
Thus in some circumstances they become as large as the $U(N)$ gluon contribution.

\begin{figure}[ptb]
\centering\subfigure{
\includegraphics[width=0.6\textwidth]{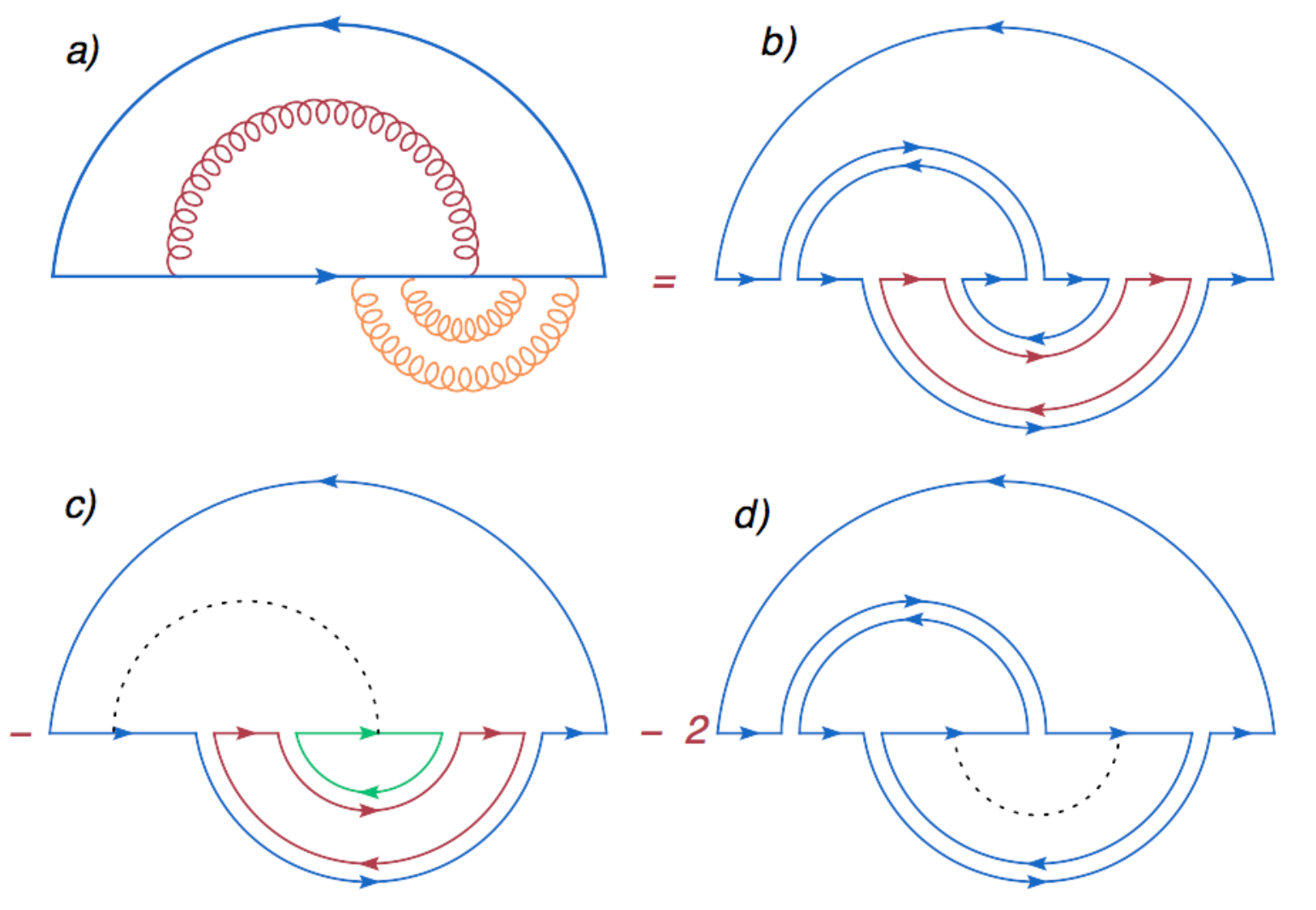} } \vspace{1cm}
\subfigure{
\includegraphics[width=0.6\textwidth]{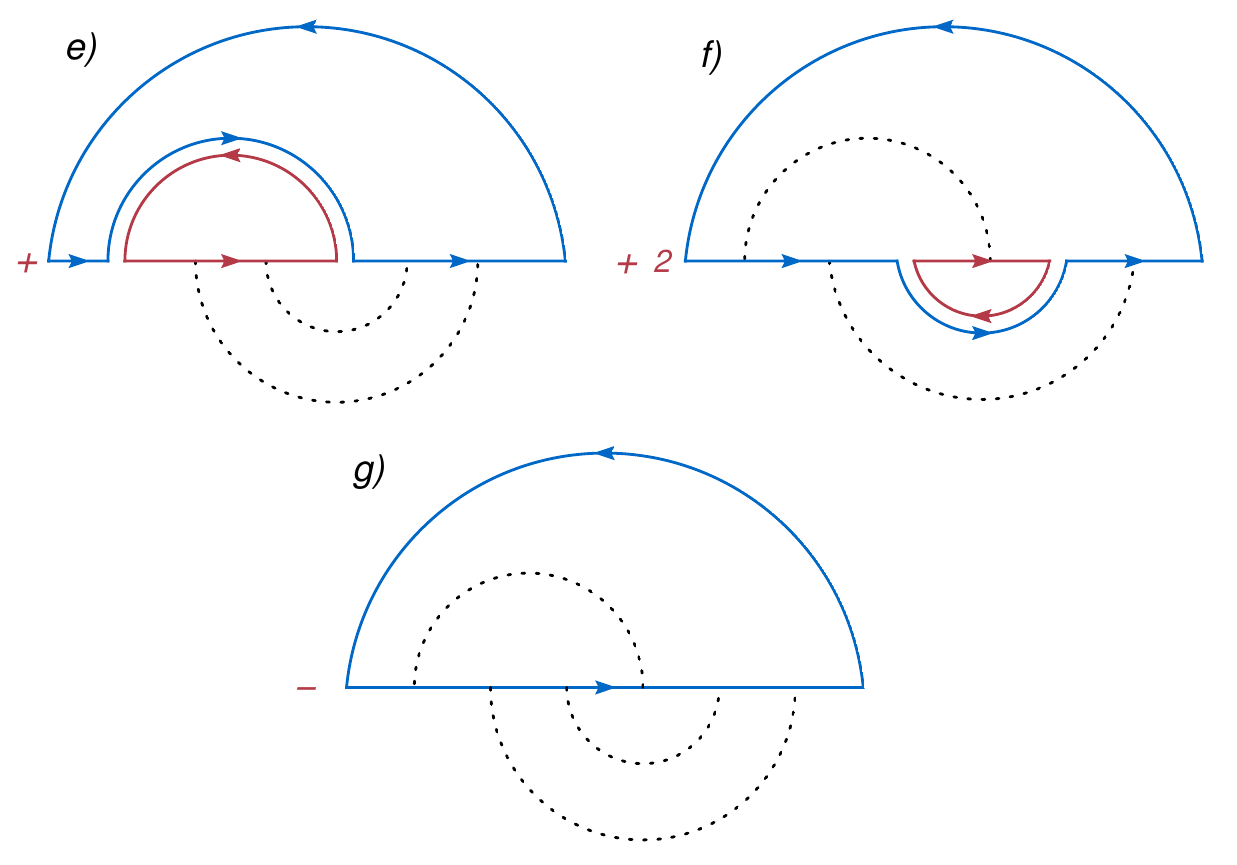}
}\caption{The double-line analysis of the diagram in Fig.
\ref{PLOT_VERT_SEN_2AD}a, where the $U(1)$ ghosts were also included.}%
\label{PLOT_SELF_EN_2A_DL}%
\end{figure}

\begin{figure}
[ptbptb]
\begin{center}
\includegraphics[width=0.6\textwidth]%
{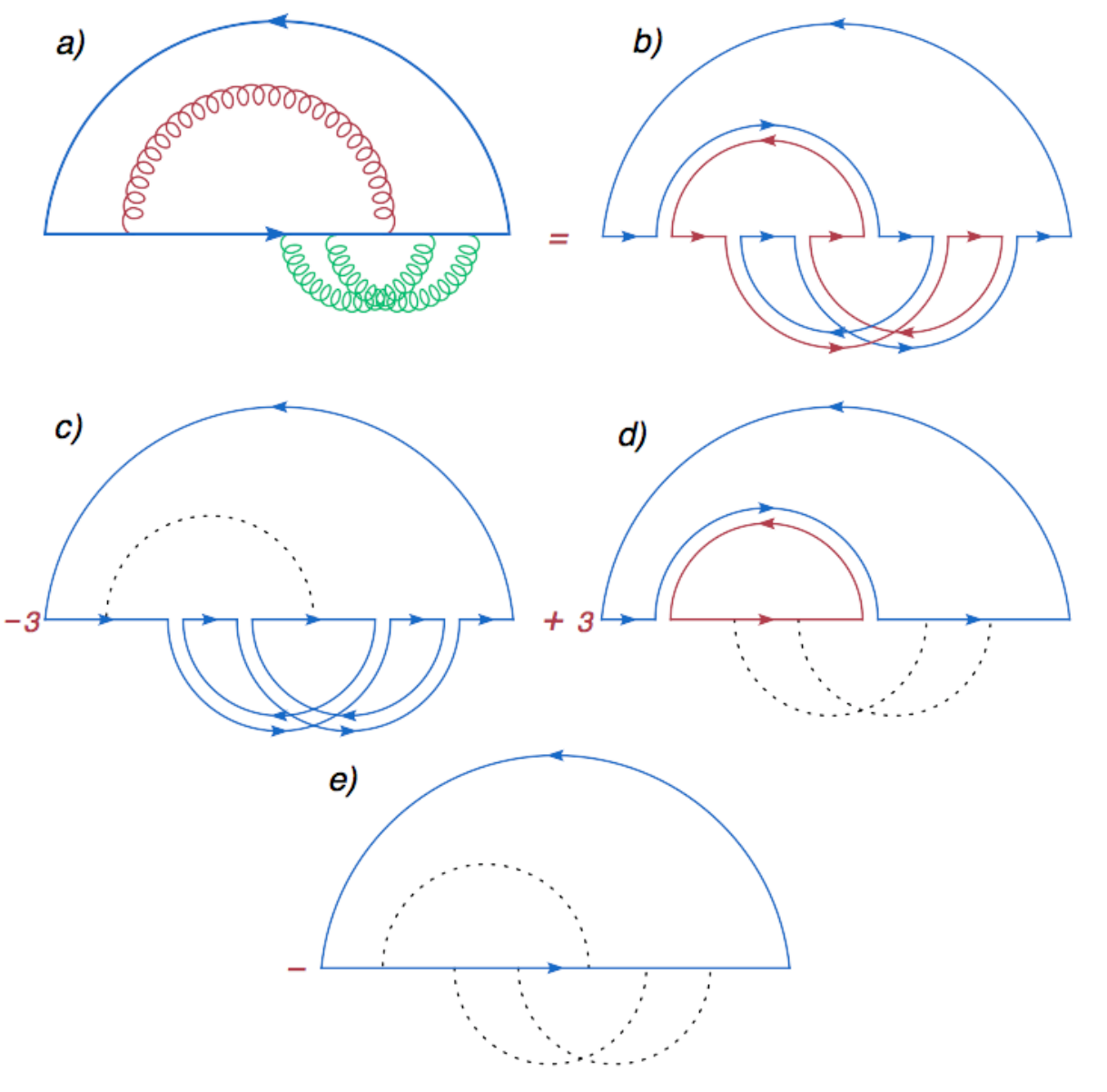}%
\caption{The double-line analysis of the diagram in Fig.
\ref{PLOT_VERT_SEN_2AD}b, where the $U(1)$ ghosts were also included.}%
\label{PLOT_SELF_EN_2D_DL}%
\end{center}
\end{figure}

The $1/N$ analysis for the vertices depicted in Fig. \ref{PLOT_VERT_SEN_2AD}
is now straightforward. The corresponding double-line decompositions are
depicted in Figs. \ref{PLOT_SELF_EN_2A_DL} and \ref{PLOT_SELF_EN_2D_DL}. We simply write down the corresponding color factors from each diagram%
\begin{gather}
\text{Cf}_{2a}^{b)}=\frac{1}{8}N^{2}\\
\text{Cf}_{2a}^{c)}=\frac{1}{8}N^{3}\times\frac{1}{N}=\frac{1}{8}%
N^{2}\nonumber\\
\text{Cf}_{2a}^{d)}=\frac{1}{8}N\times\frac{1}{N}=\frac{1}{8}\nonumber\\
\text{Cf}_{2a}^{e)}=\text{Cf}_{2a}^{f)}=\frac{1}{8}N^{2}\times\frac{1}{N^{2}%
}=\frac{1}{8}\nonumber\\
\text{Cf}_{2a}^{g)}=\frac{1}{8}N\times\frac{1}{N^{3}}=\frac{1}{8N^{2}%
}\nonumber
\end{gather}
and%
\begin{gather}
\text{Cf}_{2b}^{b)}=\frac{1}{8}N^{2}\\
\text{Cf}_{2b}^{c)}=\frac{1}{8}N\times\frac{1}{N}=\frac{1}{8}\nonumber\\
\text{Cf}_{2b}^{d)}=\frac{1}{8}N^{2}\times\frac{1}{N^{2}}=\frac{1}%
{8}\nonumber\\
\text{Cf}_{2b}^{e)}=\frac{1}{8}N\times\frac{1}{N^{3}}=\frac{1}{8N^{2}%
}\nonumber
\end{gather}


The Cf$_{2b}^{a)}$ and Cf$_{2b}^{b)}$ are easy to calculate using Eqs.
(\ref{EQ_CF_IMP}) and (\ref{EQ_CF_NPL})%
\begin{equation}
\text{Cf}_{2a}^{a)}=\text{Cf}_{2}^{\text{Pl}}Tr\left[  t^{a}t^{a}\right]
=\frac{N^{2}-1}{8N_{c}^{2}},
\end{equation}%
\begin{equation}
\text{Cf}_{2b}^{a)}=\text{Cf}_{2}^{\text{NPl}}Tr\left[  t^{a}t^{a}\right]
=\frac{N_{c}^{4}-1}{8N_{c}^{2}}.
\end{equation}

Thus we can again check that%
\begin{equation}
\text{Cf}_{2a}^{a)}=\text{Cf}_{2a}^{b)}-\text{Cf}_{2a}^{c)}-2\text{Cf}%
_{2a}^{d)}+\text{Cf}_{2a}^{e)}+2\text{Cf}_{2a}^{f)}-\text{Cf}_{2a}^{g)},
\end{equation}%
\begin{equation}
\text{Cf}_{2b}^{a)}=\text{Cf}_{2b}^{b)}-3\text{Cf}_{2b}^{c)}+3\text{Cf}%
_{2b}^{d)}-\text{Cf}_{2b}^{e)},
\end{equation}
where we skipped constructing the full $1/N$ ordering of the diagrams, as the
factor $N^{-3}$ from the gauge coupling, $g$, is common for all the diagrams.

\section{Conclusions}

 We would like to start our discussion by emphasizing that the criterion of planarity in large $N_{c}$ QCD is only valid for color-singlet diagrams with an outmost quark loop. If a non color-singlet diagram can be closed into a color-singlet diagram in more than one way, then the associated color factors will be different, thus setting the diagram at different order in the $1/N_{c}$ expansion. Moreover, if the quark-loop is not the outmost loop of the diagram, then the planarity criterion is not applicable to the given diagram that yields its $1/N_{c}$ behavior - see Ref. \cite{Manohar:1998xv} for discussion.
 
   We demonstrated that the usual double-line notation is insufficient for evaluating some non-planar diagrams, as the $SU(N_{c})$ algebra yields different results from the double-line notation. This discrepancy is caused by omitting  the $U(1)$ ghost contributions, assuming that they contribute only in subleading orders. While this is true for planar diagrams, ghost contributions must be included in order to recover the correct $1/N_{c}$ ordering of the non-planar diagrams. We derived a simple addition to the double-line notation, that provides a convenient tool for calculating both the color factors and the $1/N_{c}$ order of diagrams. The $1/N_{c}$ analysis of the quark-gluon vertex diagrams, using these new rules, completely agrees with
the straightforward calculations of the color factors using the properties of
the Gell-Mann matrices. It is remarkable that in the diagram of Fig.
\ref{PLOT_SELF_EN_2A_DL}, the leading-order contribution from $U(N)$ gluons,
Fig. \ref{PLOT_SELF_EN_2A_DL}b, is exactly canceled by a diagram including a
ghost line, Fig. \ref{PLOT_SELF_EN_2A_DL}c, so that the leading order
contribution is given by diagrams in Figs. \ref{PLOT_SELF_EN_2A_DL}%
d-\ref{PLOT_SELF_EN_2A_DL}f, containing one or more ghost lines.

\begin{acknowledgments}
This work was supported in part by DOE contract DE-AC05-06OR23177, 
under which Jefferson Science Associates, LLC,
operates Jefferson Lab. The authors would like to especially mention Jos\'e Goity for stimulating discussions and useful suggestions. HHM thanks Elizabeth Jenkins for encouraging conversations and Jerry P. Draayer for his support during the course of the work. 
\end{acknowledgments}

\bibliographystyle{apsrev}
\bibliography{Article-Color}

\end{document}